\documentclass[aps,pra,onecolumn,reprint,showpacs,superscriptaddress,amsfonts,amsmath,amssymb]{revtex4-1}

\usepackage{graphicx}
\usepackage{amssymb}
\usepackage[colorlinks=true,citecolor=hblue,linkcolor=hblue,urlcolor=hblue,pdfborder={0 0 0}]{hyperref}

\usepackage{amsfonts}
\usepackage[figuresright]{rotating}  
\usepackage{amsmath}
\usepackage{psfrag}
\usepackage{subfigure}
\usepackage{multirow}
\usepackage{tabularx}
\usepackage{textcomp}

\usepackage{verbatim}%for comments
\usepackage{amsmath}
\usepackage[colorinlistoftodos, textwidth=4cm, shadow]{todonotes}
\usepackage{xfrac}

\def\nn{\nonumber}

\def\c{\hspace{2pt}}

\renewcommand{\v}[1]{\ensuremath{\mathbf{#1}}} % for vectors
 
\newcommand{\uv}[1]{\ensuremath{\mathbf{\hat{#1}}}} % for unit vector
\let\baraccent=\= % rename builtin command \= to \baraccent

\def\beqa{\begin{eqnarray}}
\def\eeqa{\end{eqnarray}}

\newcommand{\beq}{\begin{equation} \vspace{-0em}} 
\newcommand{\eeq}{\vspace{-0.em} \end{equation}}

\newcommand{\integralb}[3]{\int\limits_{#1}^{#2} \! \mathrm{d} #3\,} 	% integral
\newcommand{\integral}[1]{\int \! \mathrm{d} #1\,}                    % integral without limits

\def\br{{\bf r}}
\def\bx{{\bf x}}
\def\bk{{\bf k}}
\def\bK{{\bf K}}
\def\bR{{\bf R}}
\def\bB{{\bf B}}
\newcommand{\dr}{^{\dagger} }

\newcommand{\beal}{\begin{align}}
\newcommand{\eeal}{\end{align}}

\newcommand{\beqs}{\begin{equation*}}
\newcommand{\eeqs}{\end{equation*}}

\newcommand{\bra}[1]{\left\langle#1\right|}  								% <...|
\newcommand{\ket}[1]{\left|#1\right\rangle}  							 	% |...>

\hypersetup{
   colorlinks=false,
   pdfborder={0 0 0},
}

\usepackage{color}
\begin{document}

\title{Ground state of a two component dipolar Fermi gas in a harmonic potential}

\author{Przemyslaw~Bienias}
\email[Corresponding author: ]{przemek@itp3.uni-stuttgart.de}
\affiliation{Center for Theoretical Physics PAN, Warsaw, Poland}
\affiliation{Institute for Theoretical Physics III, University of Stuttgart, Germany}
\affiliation{5. Physikalisches Institut, University of Stuttgart, Germany}
\author{Krzysztof~Paw{\l}owski}
\affiliation{Center for Theoretical Physics PAN, Warsaw, Poland}
\affiliation{5. Physikalisches Institut, University of Stuttgart, Germany}
\author{Tilman~Pfau}
\affiliation{5. Physikalisches Institut, University of Stuttgart, Germany}
\author{Kazimierz~Rz{a}\.zewski}
\affiliation{Center for Theoretical Physics PAN, Warsaw, Poland}
\affiliation{5. Physikalisches Institut, University of Stuttgart, Germany}

\date{\today}

\begin{abstract}
Interacting two component Fermi gases are at the heart of our understanding of macroscopic quantum phenomena like superconductivity. Changing nature of the interaction is expected to head to novel quantum phases.
Here we study the ground state of a two component fermionic gas in a harmonic potential with dipolar and contact interactions. 
Using a variational Wigner function we present the phase diagram 
of the system with equal but opposite values of the magnetic moment.  
We identify the second order phase transition from paramagnetic to ferronematic phase.

Moreover, we show the impact of the experimentally relevant magnetic field on the stability and the magnetization of the system. 
We also investigate a two component Fermi gas with
large but almost equal values of the magnetic moment to study how the interplay between contact
and dipolar interactions affects the stability properties of the mixture.
To be specific we discuss experimetally relevant parameters for ultracold $^{161}$Dy.
\end{abstract}
%03.75.Ss	Degenerate Fermi gases
%05.30.Fk	Fermion systems and electron gas (see also 71.10.-w Theories and models of many-electron systems; see also 67.10.Db Fermion degeneracy in quantum fluids)
%S	75.80.+q	Magnetomechanical and magnetoelectric effects, magnetostriction
%M	75.80.+q	Magnetomechanical effects, magnetostriction (for magnetostrictive devices, see 85.70.Ec)
%71.10.Ca	Electron gas, Fermi gas
%	31.15.E-	Density-functional theory

\maketitle

%%%%%%%%%%%%%%%%%%%%%%%%%%%%%%%%%%%%%%%%
Since the achievement of dipolar BEC \cite{Griesmaier2005} many-body physics of the ultra-cold dipolar systems attracts a lot of attention
\cite{Lahaye2009}. 
After the first sub-Doppler cooling of $^{167}$Er \cite{Berglund2007},  the fermionic dysprosium isotope $^{161}$Dy \cite{Lu2012b} was brought into quantum degeneracy 
which 
opens a new frontier for exploring strongly correlated Fermi systems. 
It  may shed some light on the properties of Quantum Liquid Crystals without unwanted solid state material complexity and disorder \cite{Kivelson1998}. 
The competition of short and long range interactions might lead to a non Fermi liquid
behavior similar to the electron ordering in an iron-based superconductor \cite{Fradkin2010b}.

Many-body studies of a polarized, one component gas in a trap revealed that its ground state has only uniaxial symmetry in position space~\cite{Goral2001}. 
Moreover, the exchange energy \cite{gross1995} leads to the stretch of the Fermi surface along polarization axis and changes the stability properties~\cite{Miyakawa2008a,Zhang2009b}. 
Such a deformation can be imaged by time-of-flight technique~\cite{He2008,Kohl2005a,Lima2010b,Lima2010}. 
Finally, breaking of uniaxial symmetry for sufficiently strong interaction is possible, what leads to a biaxial phase~\cite{Fregoso2009a}.

Although close to the spin-\sfrac{1}{2} electron case ground state properties of two component system are much less explored. 
For a homogeneous 3D gas the existence of the ferronematic phase was found~\cite{Fregoso2009}.  
{Namely,} for a strong enough contact interaction the ground state has nonzero magnetization and the Fermi surfaces have only uniaxial symmetry. 
The transition from a paramagnetic phase to a ferronematic one is possible by increasing only the strength of the dipolar interaction.
Interestingly, partial magnetization occurs for suitable dipolar and contact coupling constants. 
For the 2D system in a box an inhomogeneous external magnetic field was taken into account~\cite{Fang2011a}.

But till now the investigation of the ground state properties of the 3D gas in the trap was lacking. 
The purpose of this paper
%Letter 
is to present the first study of the two component fermionic system in the 3D harmonic trap with long-range dipolar and short-range isotropic interactions.
Such a study is especially relevant for the upcoming experiments with fermionic Dy or Er.
The most important is our finding of the second order phase transition from paramagnetic to partially magnetized nematic phase.

%%%%%%%%%%%%%%%%%%%%%%%%%%%%%%%%%%%%%%%%
We consider fermionic atoms of mass $m$ which can be in two hyperfine states (denoted 1 and 2), having magnetic moment $\vec{\mu}= \mu \vec{\sigma}$, where $\vec{\sigma}$ are spin-$\sfrac{1}{2}$ Pauli matrices. The Hamiltonian of the gas in the harmonic potential (with a frequency $\omega$) reads:
\beqa
\hat{H} &=& \integral{^3 x} \psi_i^{\dagger}(\v{x})
\left(-\frac{\hbar^2\nabla^2}{2m} + \frac{m \omega^2 x^2}{2} \right)\psi_i(\v{x}) 
+\nn \\
&& \frac{1}{2} \integral{^3x}  \mathrm{d}{^3 x'}  \psi_i^{\dagger}(\v{x})\psi_j^{\dagger}(\v{x}')
V_{ii'jj'}(\v{x},\v{x}')\psi_{j'}(\v{x}')\psi_{i'}(\v{x}), \nn
\eeqa
where the fields $\psi_{i}(\v{x})$ destroy fermions in a spin state with $z$-component 
$i=1,2$ at position~$\v{x}$. The fields $\psi_i(\v{x})$ satisfy standard fermionic anticommutation relations. We use the convention that repeated indices are implicitly summed over.
The interparticle potential including dipolar and contact interactions has the form:
\beqa
V_{ii'jj'}(\v{x},\v{x}') &=&
\frac{3}{4\pi} \frac{g_d}{r^3} \sigma^{l}_{ii'} ( \delta_{lm} - 3 \uv{r}_l \uv{r}_m )\sigma^{m}_{jj'} \nn \\
 && \hspace{20pt}+\c g_c\c\delta_{ii'}\delta_{jj'}\delta(\v{r}), 
 \label{eqn:potential}
\eeqa
where $g_c$ is a coupling strength of the contact interaction, $g_d=\mu_0 \mu^2/3 $ is the dipolar interaction coupling,  $\v{r}\equiv (\v{x}-\v{x}')$ and $\uv{r}$ is a unit
 vector in the direction of $\v{r}$. The Fourier transform 
of the two body interaction is $\tilde{V}_{ii\rq{}jj\rq{}} = g_{d}\sigma^{l}_{ii'}(3 
\uv{q}_{l}\uv{q}_{m} -\delta_{lm})\sigma^{m}_{jj'} + g_c \delta_{ii'}
\delta_{jj'}$.

The total energy can be expressed as a functional of Wigner functions $f_{ij}(\bx,\bk)$ \cite{Fang2011a,Zhang2009b}: 
\beqa
E[\vec{f}] &=& \integral{^3x}\integral{^3k}
\left(-\frac{\hbar^2\nabla^2}{2m} + \frac{m \omega^2 x^2}{2} \right) f_{ii}(\bx,\bk) + \nn \\
&
+& \frac{1}{2(2\pi)^6} \integral{^3\bx } \mathrm{d}{^3\bk } \mathrm{d}{^3\bx\rq{}} \mathrm{d}{^3 \bk\rq{}} 
f_{ii'}(\bx,\bk) f_{jj\rq{}}(\bx\rq{},\bk\rq{}) \nn\\ 
&\times & V_{ii\rq{}jj\rq{}}(\bx-\bx\rq{})
- \frac{1}{2(2\pi)^6}  \integral{^3\bR } \mathrm{d}{^3\bk } \mathrm{d}{^3 \bk\rq{}} \nn\\&\times&
f_{ii'}(\bR,\bk) f_{jj\rq{}}(\bR,\bk\rq{}) 
\tilde{V}_{ii \rq{}jj \rq{}}(\bk-\bk\rq{})
\label{eqn:totalFun},
\eeqa
where $\bR=\frac{\bx+\bx'}{2}$, the second and the third term are called direct and exchange energy respectively.

We propose the variational Wigner function diagonal in the spin space ($i=j$ in $f_{ij}$) which enables deformations in momentum and position spaces (parameters $\alpha$ and $\beta$ respectively) as well as compression in the position space ($\lambda$): 
\begin{align}
f_{ii}(\br,\bk)=
\frac{1}{\gamma^6} 
\exp
  \left[-
\frac{
   (k_x^2+k_y^2)/\alpha_i+\alpha_i^2k_z^2+ 
    }{\gamma^2K_i^2}\right]\nn\\
\times
\exp
  \left[-
\frac{
  \lambda_i^2
	\left[
	  \beta_i(x^2+y^2)+z^2/ \beta_i^2
	\right]
    }{\gamma^2K_i^2}\right].
\label{eqn:wignerGauss}
\end{align}
Such a choice is motivated by the anisotropic nature of the dipole-dipole interaction which leads to breaking of a spherical symmetry in position and momentum spaces.
In this paper we use oscillator units for length and energy: $l_{osc}=\sqrt{\hbar/m\omega}$ and $\hbar \omega$ respectively.

Wigner function \eqref{eqn:wignerGauss} for 
$K_i=(2\lambda_i)^{1/2}N_i^{1/6}$
fulfills the constraint $N_i=\integral{^3x}n_{ii}(\bx). $
The density distributions in position and momentum spaces are given respectively by \cite{Miyakawa2008a}:
\begin{align}
n_{ii}(\bx)=\frac{\sqrt{N_i}}{\gamma^3} \left(\frac{ \lambda_i}{2\pi}\right)^{3/2}\exp\left[-\lambda_i\frac{\beta_i(x^2+y^2)+\beta_i^{-2}z^2}{2 N_i^{1/3} \gamma^2}\right],
\label{eq:densReSpace}\\
\tilde{n}_{ii}(\bk)=\frac{\sqrt{N_i}}{\gamma^3} \frac{ 1}{\left(2\pi\lambda_i\right)^{3/2}}\exp\left[-\frac{\alpha_i^{-1}(k_x^2+k_y^2)+\alpha_i^{2}k_z^2}{2\lambda_i N_i^{1/3} \gamma^2 }\right].\nn
\end{align}
Under the Gaussian ansatz \eqref{eqn:wignerGauss} all terms in the energy functional can be evaluated analytically. The kinetic and trap energies are given by:
\begin{align}
E_{kin}^i=N_i^{4/3}\gamma^2 \lambda_i\frac{\alpha_i^{-2}+2\alpha_i}{2},
\label{eqn:kin}\\
E_{tr}^i=N_i^{4/3} \gamma^2 \frac{\beta_i^{2}+2/\beta_i}{2\lambda_i}.
\label{eqn:trap}
\end{align}

First, let us concentrate on the fully polarized one component gas. 
Then, only one Wigner function is needed: $f(\bx,\bk)$ (we omit indices $i,j$) and the potential Eq. \eqref{eqn:potential} has the form: $\frac{3}{4\pi} \frac{g_d}{r^3} ( \delta_{lm} - 3 \uv{r}_l \uv{r}_m )$.
 Contact energy vanishes due to the interplay between direct and exchange energies\cite{Goral2001}.
Finally, the direct and exchange energies for one component are \cite{Miyakawa2008a}:
\begin{align}
E_{dir}=-g_{d}N^{3/2}\gamma^{-3}2^{-4}\pi^{-3/2}(\beta\lambda)^{3/2}g_{dip}(\beta^3),
\label{eqn:dirEn}
\\
E_{ex}=g_{d}N^{3/2}\gamma^{-3}
2^{-4}\pi^{-3/2} (\alpha\lambda)^{3/2}
g_{dip}(\alpha^3),
\label{eqn:exEn}
\end{align}
where:
$
g_{dip}(\kappa)=
\integralb{0}{1}{u}
({1-3u^2})
{(1-(1-\kappa)u^2)^{-3/2}}=
{(-3 \sqrt{\kappa-1}+(2+\kappa) \arctan\left[\sqrt{\kappa-1}\right])}{( \kappa-1)^{-3/2}}.
$
The total energy is equal to the sum of terms \eqref{eqn:kin}, \eqref{eqn:trap}, \eqref{eqn:dirEn} and \eqref{eqn:exEn}:
$E= E_{kin}+ E_{trap} + E_{dir}+ E_{ex} \label{eqn:enFun} $.

%%%%%%%%%%%%%%%%%%%%%%%%%%%%%%%%%%%%%%%%
For the one component gas Zhang et al. have shown \cite{Zhang2009b} that a $\theta$-Heaviside function ansatz on a Wigner function \cite{Miyakawa2008a}
$
f(\bx,\bk)=\Theta 
\bigg[ k_F^2-\frac{1}{\alpha}
\left(k_x^2+k_y^2\right)-\alpha^2k_z^2+\nn\\
-\frac{\lambda^2}{l^4_{osc}}
\left[\beta
\left(x^2+y^2\right)+\frac{1}{\beta^2}z^2 \right] 
\bigg],
$
describes very well the system except in the region close to the collapse.  
For the $\theta$-Heaviside Wigner function 
kinetic and trap energies are: 
\begin{align}
E_{kin}=c_1 N^{4/3} \lambda (\alpha^{-2}+2\alpha),
\label{eqn:kinGauss}
\\
 E_{tr}=N^{4/3}c_1 (\beta^{2}+2/\beta)/\lambda, 
 \label{eqn:trapGauss}
\end{align}
where $c_1=3^{1/3}/2^{8/3}$.

\begin{figure}[h!]
\includegraphics[width=0.37\textwidth]{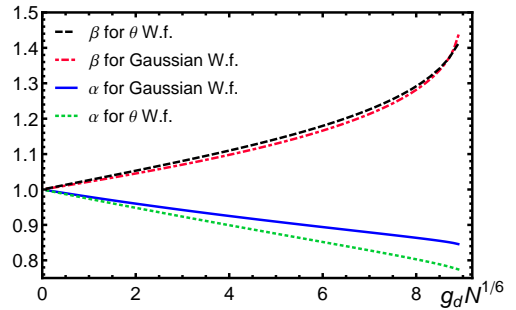}\caption{\label{fig:fig1}
Comparison of the deformation of the cloud in momentum ($\alpha$) and position ($\beta$) spaces between Gaussian and $\theta$ -Wigner function (W.f.) for the $g_{d}$ in the stable regime. Notice very good agreement for stable range of $g_{d}N^{1/6}$.}
\end{figure}

We choose parameter $\gamma$ such that our Gaussian ansatz gives the same kinetic and potential energy as $\theta$-Heaviside function ansatz for the same deformations and compression of the cloud. From \eqref{eqn:kin}, \eqref{eqn:trap}, \eqref{eqn:kinGauss}, \eqref{eqn:trapGauss} we get that   $\gamma=\sqrt{2 c_1}$. 
Next, using Nelder Mead method, we find a local minimum of the energy functional which emerges from the adiabatic switching of dipolar and contact interactions.
We show in Fig.~\ref{fig:fig1} the comparison of deformations in position and momentum spaces between the $\theta$-Heaviside and the Gaussian Wigner functions. 
We see a qualitative agreement between both ansatzes.
Notice a good agreement of the range of stability between both ansatzes.
The physics of two component gas is considerably richer than that of one component gas. 
First, the contact interaction between two different components is present. For Gaussian Wigner function \eqref{eqn:wignerGauss} it has the form:
\beq
E_{c}=g_c\frac{\lambda_1^{3/2}\lambda_2^{3/2}\sqrt{N_1 N_2}\gamma^{-3}(2\pi)^{-3/2}}{
\sqrt{\frac{\lambda_1}{\beta_1^2N_1^{1/3}} + \frac{\lambda_2}{\beta_2^2N_2^{1/3}} }
 \left( \frac{\lambda_1 \beta_1}{N_1^{1/3}} + \frac{\lambda_2 \beta_2}{N_2^{1/3}}\right)
}.
\label{eq:contact}
\eeq
It leads to the conventional Stoner transition\cite{Stoner1947} and simultaneously stabilizes each component against deformations in momentum and position spaces.

Second, $V_{1221}$ is nonzero, so from Eq. \eqref{eqn:totalFun} we see that in general terms like $f_{12}(\bx,\bk)$ (called correlations between components) might appear. 
However, the ground state of an ideal gas is a product  $\ket{N_1}\ket{N_2}$ so the intercomponent correlation function $\bra{N_1}\bra{N_2} \psi_1^{\dr}\psi_2\ket{N_1}\ket{N_2}$ vanishes and $f_{12}(\bx,\bk)$ as well.

From a continuity argument  for  $f_{12}$ as a function of $g_d$ we can neglect for small $g_d$ the contribution of intercomponent correlations. 
Moreover, for unstable and fully magnetized ferronematic phase, properties of the system depend only on one component so correlations are not present.
In further analysis we  neglect correlations making the problem numerically tractable but paying the price of presenting more  qualitative than quantitative results.

The third, dipolar energies for two components emerging from $V_{1212}$ and $V_{2121}$ {have much more complicated form} than for a single component. The direct energy between the two components is:
\beqa
E_{dir}^{12}=
-\frac{g_{d}N_1N_2}{2}\gamma^{-3}(2\pi)^{-3/2}  
\frac{g_{dip}(\kappa_{dir})}{\left(\frac{N_1^{1/3}}{\beta_1\lambda_1}+\frac{N_2^{1/3}}{\beta_2\lambda_2} \right)^{3/2}},
\label{eq:2CompEDir}
\eeqa
where: 
$
\kappa_{dir}=
{(
\frac{N_1^{1/3}\beta_1^2}{\lambda_1} + \frac{N_2^{1/3}\beta_2^2}{\lambda_2}
)}/{(
\frac{N_1^{1/3}}{\beta_1\lambda_1}+\frac{N_2^{1/3}}{\beta_2\lambda_2}
)}.
$
While the exchange energy between the two components is equal to:
\beqa
E_{ex}^{12}&=&
\frac{\frac{1}{2}g_{d}\gamma^{-3}(2\pi)^{-3/2}}{
\sqrt{\frac{\lambda_1}{\beta_1^2 N_1^{1/3}}+\frac{\lambda_2}{\beta_2^2 N_2^{1/3}}}
\sqrt{\frac{\alpha_1^2}{\lambda_1 N_1^{1/3}}+\frac{\alpha_2^2}{\lambda_2 N_2^{1/3}}}
}\nn\\
&\times&\frac{(\alpha_1\lambda_1N_1^{1/3}+\alpha_2\lambda_2 N_2^{1/3})^{3/2} g_{dip}({\kappa_{ex}})}{
\left(
\frac{\beta_1\lambda_1 }{N_1^{1/3}}+\frac{\beta_2\lambda_2 }{N_2^{1/3}}
\right)\times
\left(\frac{1}{\alpha_1\lambda_1 N_1^{1/3}}+\frac{1}{\alpha_2\lambda_2 N_2^{1/3}}\right)}
,
\label{eq:2CompEEx}
\eeqa
where $\kappa_{ex}=
{(\alpha_1 \lambda_1 N_1^{1/3}+\alpha_2\lambda_2 N_2^{1/3})}/{(\frac{\lambda_1 N_1^{1/3}}{\alpha_1^2}+\frac{\lambda_2 N_2^{1/3}}{\alpha_2^2})}.
\label{eq:2CompKappaEx}
$
For the derivation see Supplemental material. 
The total energy is equal to 
\begin{align} E^1_{kin}+E^1_{pot}+E^2_{kin}+E^2_{pot}+2(E_{dir}^{12}+E_{ex}^{12}) +\nn\\+E^1_{dir} + E^1_{ex} +E^2_{dir} + E^2_{ex}+E_c,
\label{eq:totalEn}
\end{align}
where single indexes $1$ or $2$ denotes component for which we take the kinetic, trap, direct and exchange energies, Eq. \eqref{eqn:kin}, \eqref{eqn:trap}, \eqref{eqn:dirEn} and \eqref{eqn:exEn} accordingly .
\begin{figure}[t!]
\subfigure{\includegraphics[width=0.37\textwidth]{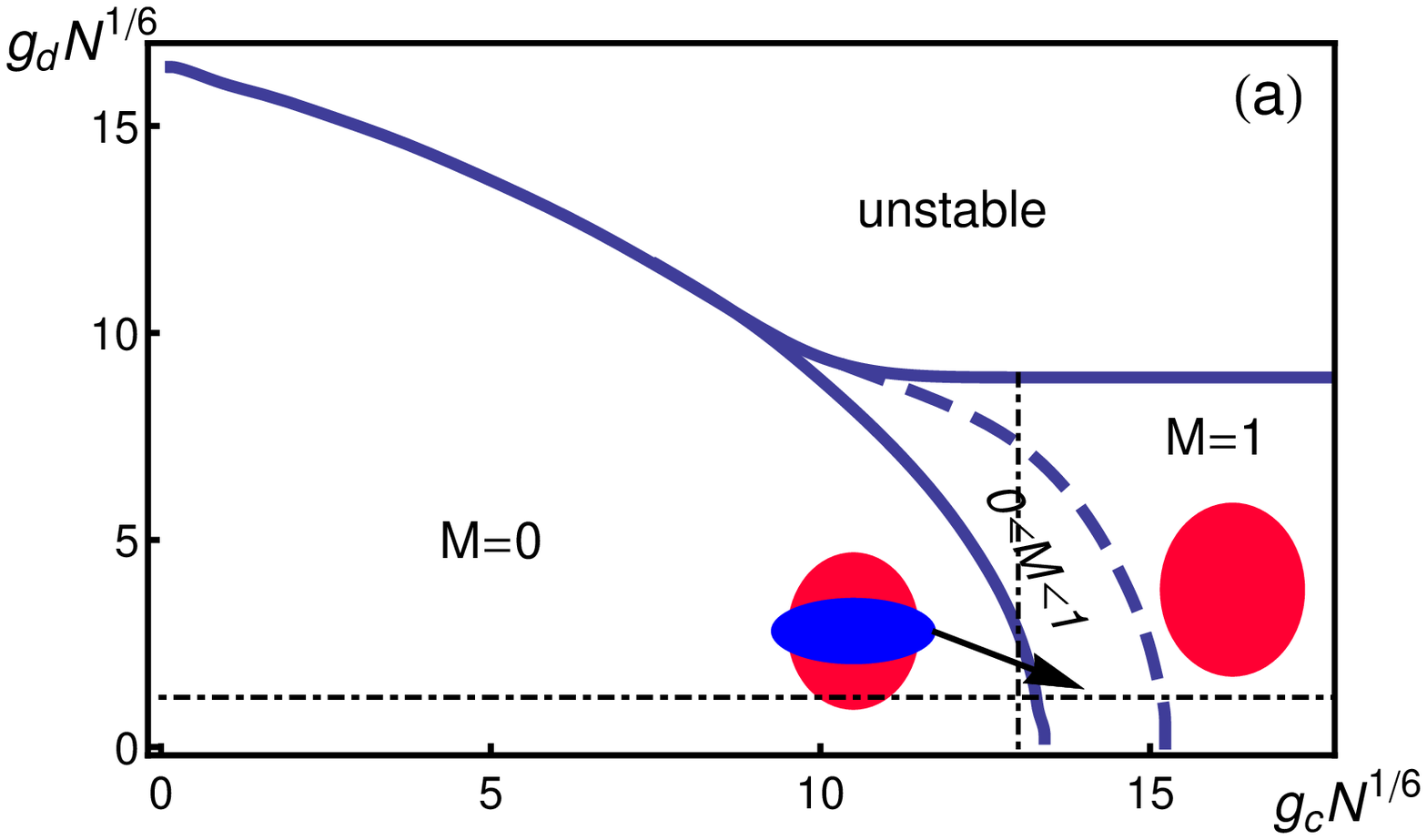}}\\
\vspace{-1.3em}
\subfigure{\includegraphics[width=0.49\textwidth]{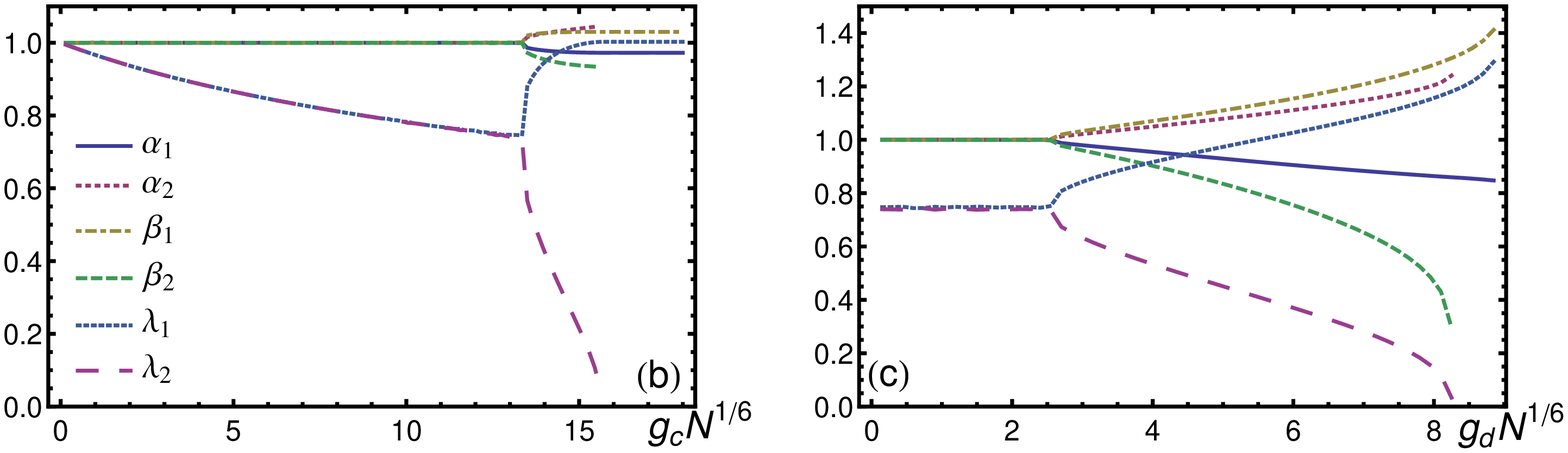}}
\caption{
 (a) Phase diagram for two component gas. 
Three regimes of different magnetization are visible. For $M=0$ size and shape of components are like for noninteracting gas. 
For $0<M<1$ Fermi surface of bigger component has the prolate shape (red) whereas for smaller one the oblate (blue). 
Thick, dashed line presents smooth crossover to the phase with $M=1$ where shape of the Fermi surface is prolate. 
The transition to the unstable regime is possible from ferronematic but also from unmagnetized phase.
(b-c) Deformations in momentum and position spaces ($\alpha_i$ and $\beta_i$) as well as compression in the position space ($\lambda_i$) for constant $g_dN^{-1/6}=1.2$ and $g_cN^{-1/6}=13$ respectively (dash-dotted lines in (a)). 
In paramagnetic phase, due to the contact interaction, both components are getting bigger in position space with growing $g_c$. 
In ferronematic regime smaller component (index 2) still gets bigger while for constant $g_d$ parameter $\lambda_1$ goes to 1 what means equal occupation of the position and momentum phase spaces.  For constant $g_c$ we see that the collapse is due to the higher occupation of the momentum rather than position phase space.
}
\label{fig:phDiag20130325Squeezed}
\end{figure}
Fig. \ref{fig:phDiag20130325Squeezed} shows the magnetization ($M=|N_{1}-N_{2}|/N$) and the stability of the system as a function of the dipolar and the contact coupling constants.
In the area of $M=0$ (paramagnetic phase) both components have the same spherical shape. 
In the area where $0<M<1$ we have a ferronematic phase with partial magnetization. 
Of course the ground state is doubly degenerate. Either component could be a dominant one.
The Fermi surface for the component with a higher occupation is prolate whereas for a lower occupation is oblate. 
For $M=1$ only one component is occupied and the Fermi surface is prolate. 
The unstable phase has a boundary with ferronematic (first order phase transition) and unmagnetized phase whereas for gas in a box only the transition from ferronematic to unstable regime is possible \cite{Fregoso2009}. 
On the other hand, in the trap, no direct transition from paramagnetic to ferronematic phase with M=1 is possible. 
The phase transition from paramagnetic to ferronematic with $M\neq 1$ is of the second order.
Moreover, the partially magnetized ferronematic phase is much larger than for the box and extends up to the unstable regime.
Using $\pi$ polarized light it is possible, due to the quadratic dependence of the AC Stark on magnetic quantum number, to prepare a system of $^{161}$Dy atoms in two extreme $m_{1,2}=\pm\sfrac{21}{2}$ states. In this case  $N=10^6$ atoms  and $\omega=2\pi\times 100 $Hz corresponds to $g_dN^{1/6}\approx 1$ in Fig. \ref{fig:phDiag20130325Squeezed}. By tuning the number of atoms 
%and using yet uknown Feshbach resonances 
it is possible to investigate the large part of the phase diagram.

The above analysis (as well as other presented in the literature) assumes that external magnetic field $\bB$ is so small that it can be neglected. 
But in real cold atoms experiment the magnetic field may be controlled up to $10 \mu G$ \cite{Pasquiou2011b}. One might ask a question what impact on the phase diagram has the $\bB$ field? 
\begin{figure}[t!]
\subfigure%[$ g_BN^{-1/3}=0.0014$]
{\includegraphics[width=0.23\textwidth]{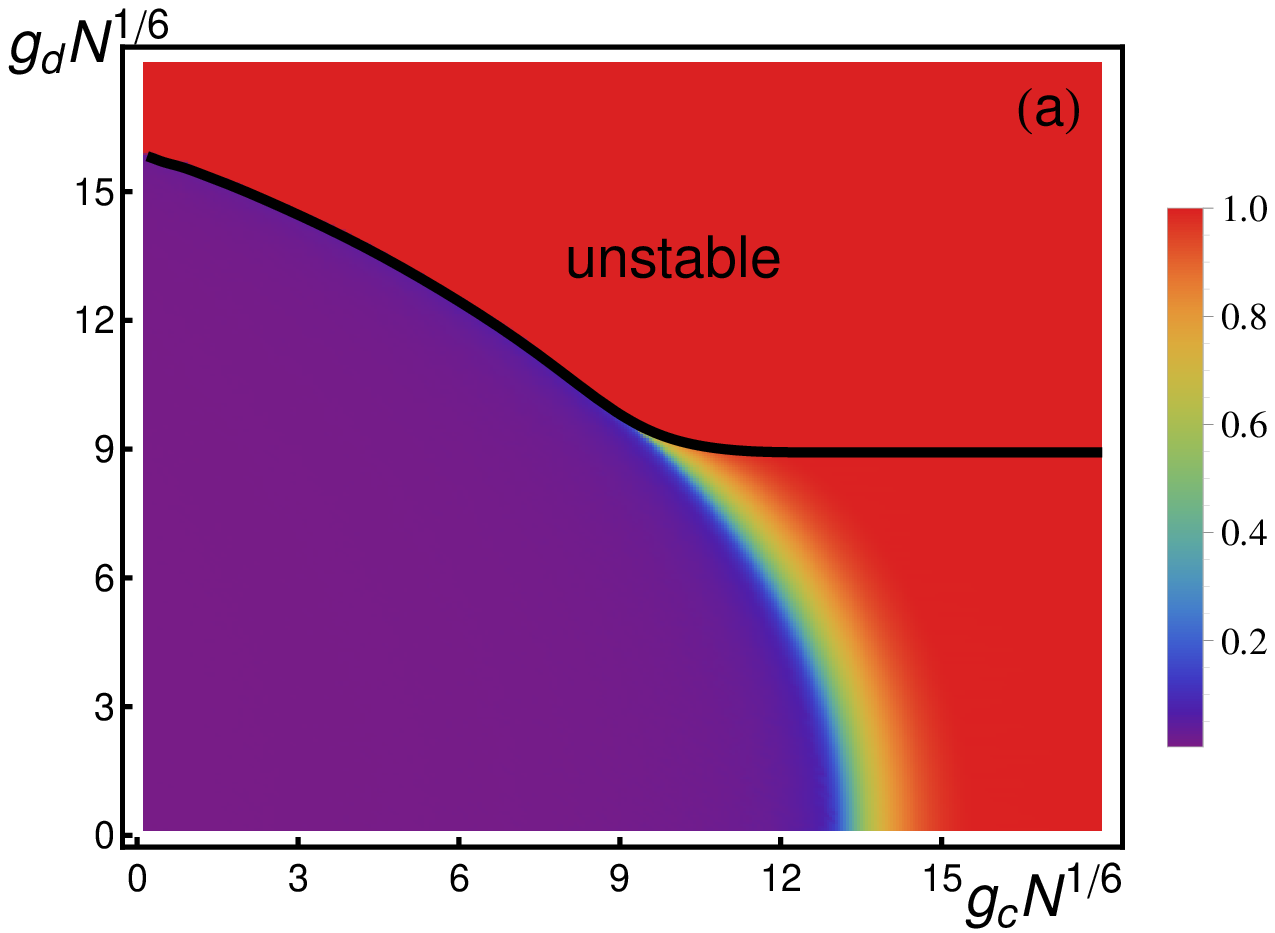}}
\subfigure%[$g_BN^{-1/3}=0.14$]
{\includegraphics[width=0.23\textwidth]{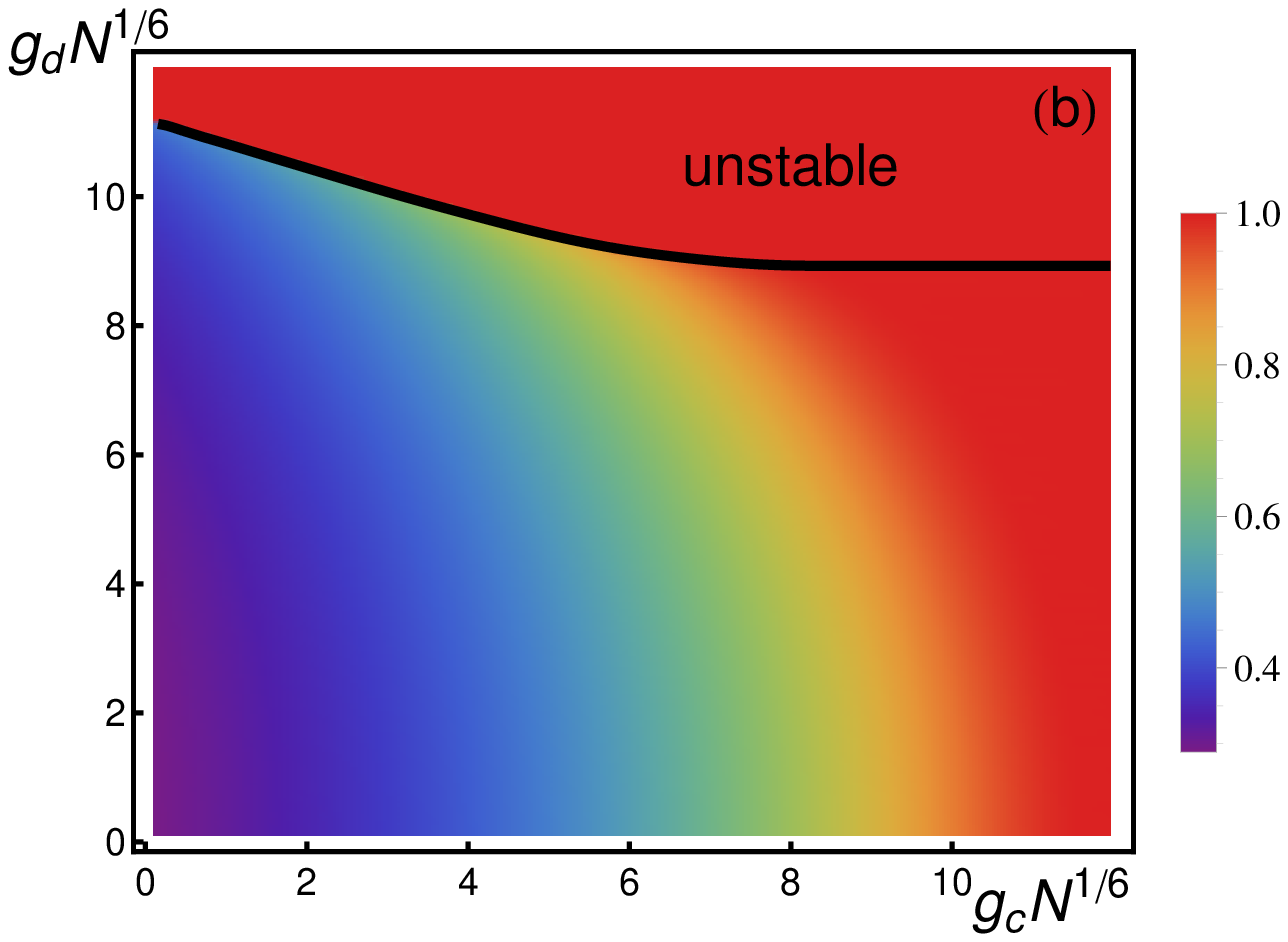}}\\
\vspace{-1.1em}
\subfigure%[$g_BN^{-1/3}=0.28$]
{\includegraphics[width=0.23\textwidth]{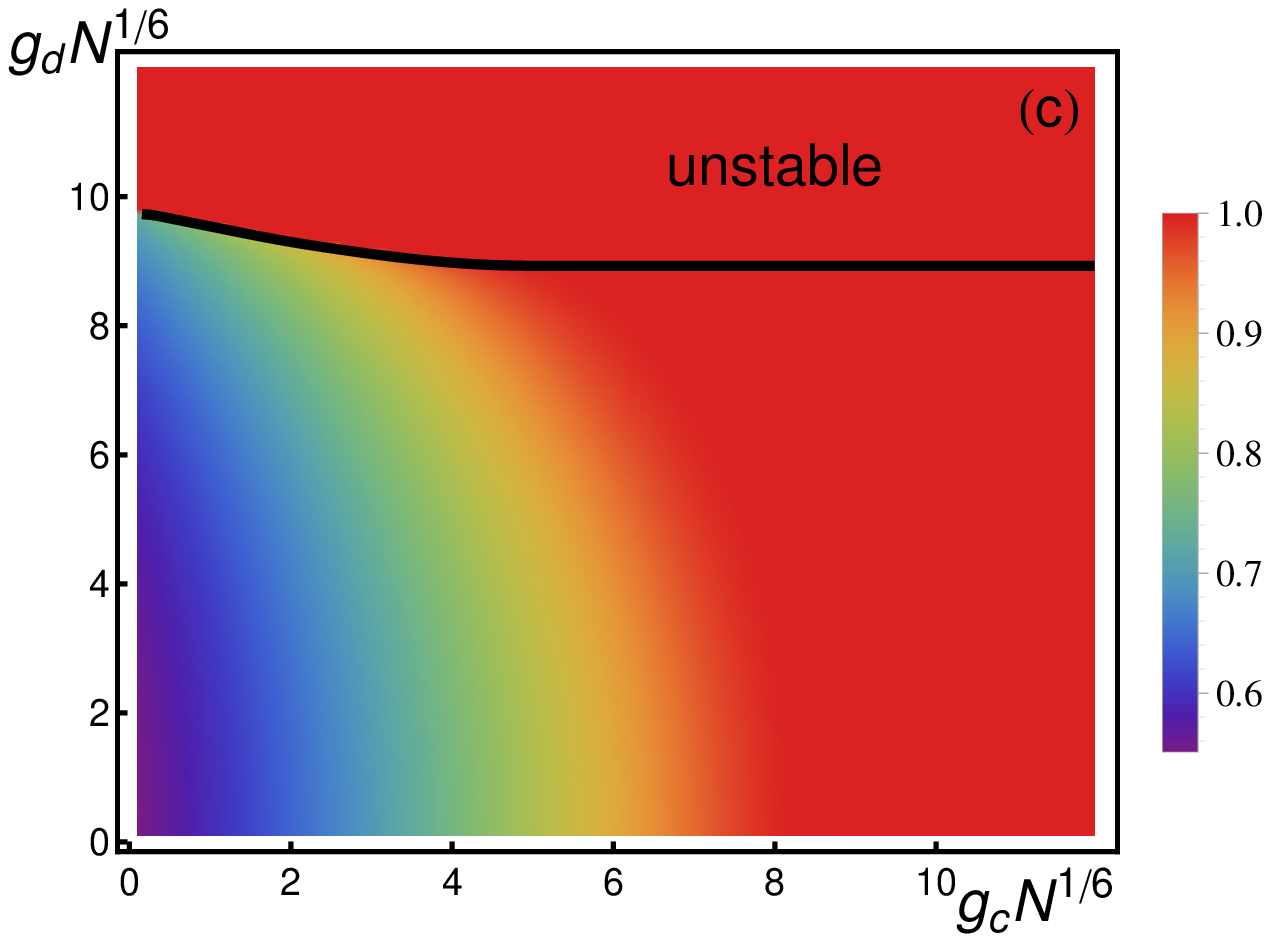}}
\subfigure%[$g_BN^{-1/3}=0.56$]
{\includegraphics[width=0.23\textwidth]{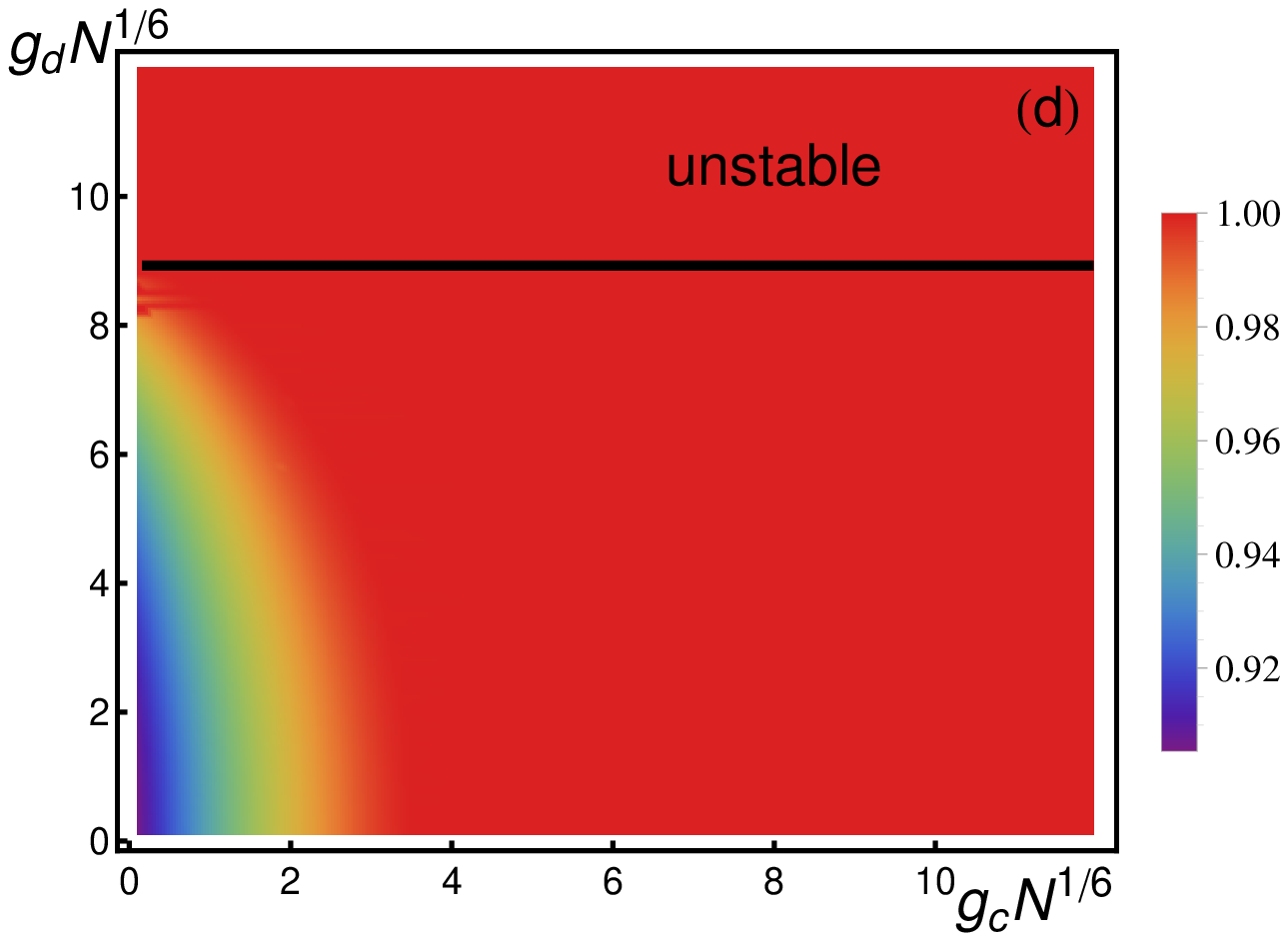}}
\caption{\label{fig:bField} (a-d) Phase diagrams showing magnetization and unstable regime for  
several values of $g_B N^{-1/3}$ 0.0014, 0.14, 0.28 and 0.56 respectively. For nonzero magnetic field we no longer have paramagnetic phase.
Black solid line shows the range of the unstable regime. Notice that the boundary between unstable and fully polarized ferronematic phase does not depend on the strength of magnetic field. For $N=10^6$ and $\omega=2\pi\times 100$Hz diagrams correspond to $B=0.01, 1, 2, 4$ mG.
}
\end{figure}
In the presence of a homogeneous magnetic field, the energy functional \eqref{eqn:totalFun} is changed by addition of the term: $-g_B(N_{1}-N_{2}) $, where we defined $g_B$ as equal to $\mu B$ and used the fact that the magnetic field is parallel to a quantization axis of the spin-$\sfrac{1}{2}$. 
Results for different $g_B N^{-1/3}$ are presented in Fig.~\ref{fig:bField}. 
 For nonzero magnetic field we no longer have a paramagnetic phase and magnetization of the system continuously changes with $g_d$ and $g_c$.
For the noninteracting gas we had spherical Fermi surfaces. 
They get more and more prolate ($1$st component) and oblate ($2$nd) with growing dipolar interaction up to the unstable phase.
Of course now, the magnetic field favors the component with magnetic moment parallel to the magnetic field.
Black solid line shows the range of the unstable regime. 
Notice that boundary between the unstable and fully polarized ferronematic phases does not depend on the strength of the magnetic field. Moreover, values of the magnetization for $g_BN^{-1/3}=0$ and $g_BN^{-1/3}=0.0014$ differ by less than 5\%. 
\begin{figure}[t!]
\subfigure{\includegraphics[width=0.37\textwidth]{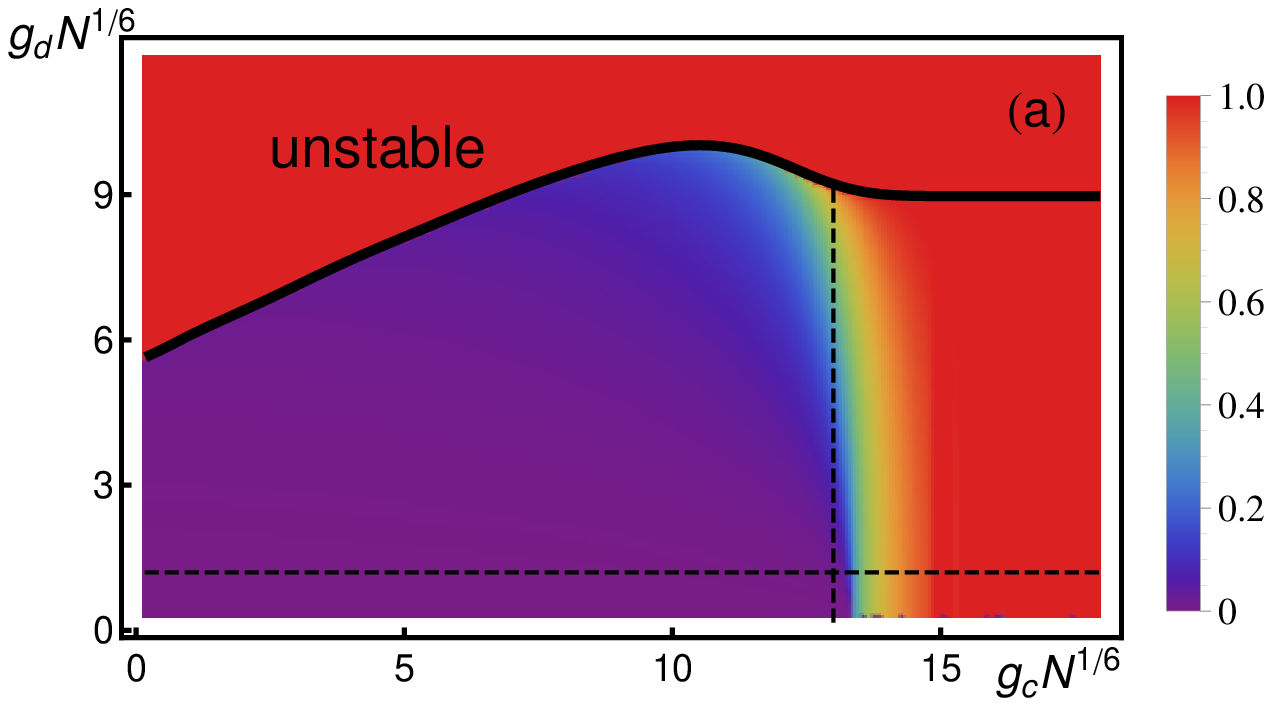}}\\
\vspace{-1.em}
\subfigure{\includegraphics[width=0.49\textwidth]{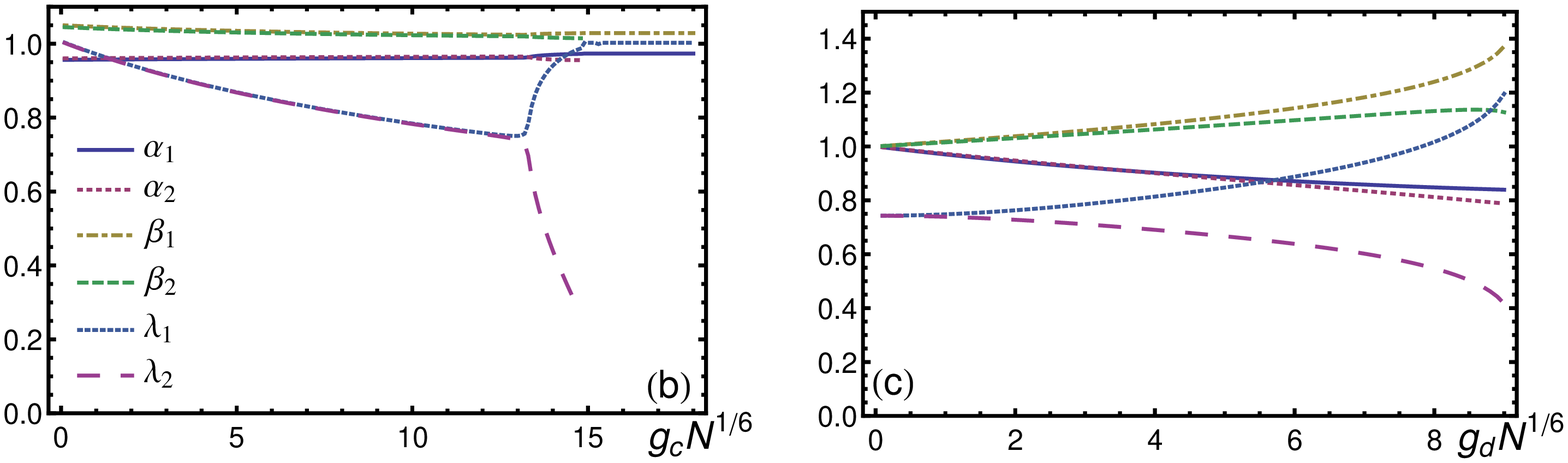}}
\caption{ (a) Diagram showing $\frac{N_1-N_2}{N}$ for two component gas of $m_1=\sfrac{21}{2}$ and $m_2=\sfrac{19}{2}$ for different number of atoms and $\omega=2\pi\times 100$Hz. 
 (b-c) Deformations in momentum and position spaces ($\alpha_i$ and $\beta_i$) as well as compression in the position space ($\lambda_i$) for constant $g_dN^{-1/6}=1.2$ and $g_cN^{-1/6}=13$ respectively (dashed lines in (a)).
In whole range $N_1>N_2$ and both components, whenever exist, are prolate in momentum and position spaces (more elongated is the first one). The boundary of the stability of two components against collapse is strongly changed due to the interplay between contact and dipolar interactions. 
}
\label{fig:20130325DnCloseSqueezed}
\end{figure}

The highest possible magnetic moment from all elements of ${^{161}}$Dy (10$\mu_B$) \cite{Lu2012b} can be exploited in one more interesting system, namely in two component gas of $m_1=\sfrac{21}{2}$ and $m_2=\sfrac{19}{2}$. 
Using the magnetic quantum number dependent AC Stark shift  and the Zeeman shift it can be prepared in a degenerate state (without dipolar interaction).  
The energy functional in analogy to Eq. \eqref{eq:totalEn} has the form:
$E^1_{kin}+E^1_{pot}+E^2_{kin}+E^2_{pot}+E_c-2\frac{19}{21}(E_{dir}^{12}+E_{ex}^{12}) +E^1_{dir} + E^1_{ex} +(\frac{19}{21})^2(E^2_{dir} + E^2_{ex})$ what comes from replacing Pauli matrix $\sigma_z$ in Eq. \eqref{eqn:potential} by the matrix $\begin{pmatrix}
1&0\\
0&19/21
\end{pmatrix}$.
A corresponding phase diagram is presented in Fig. \ref{fig:20130325DnCloseSqueezed}. 
Whereas in the case of $m_1=-m_2$ components we had large area of $N_1=N_2$ for this system we have $N_1>N_2$ for $g_d>0$ because the dipolar interaction is stronger for the 1st component. 
Both components are prolate in momentum and position spaces but the deformation is larger for the 1st one. 
The instability boundary of a one component gas
is the same as for $m_1=m_2$
because it depends only on the larger magnetic moment. 
The most fascinating result is a nontrivial direct transition from the two component to an unstable system. Due to the interplay between the dipolar and contact interactions the last one is stabilizing the system more up to some critical value of $g_cN^{1/6}$.   

%Summarizing: W
In conclusion we have presented the first study of the ground state of the two component fermionic gas in a harmonic potential with dipolar and contact interactions.
To be specific we have chosen parameters for the fermionic isotope $^{161}$Dy in the 
 $m_{1,2}=\pm\sfrac{21}{2}$ state where we have identified regimes of unstable, paramagnetic and ferronematic phases and transitions between them. 
We also showed that experimentally accessible control over the magnetic field enables observation of the sharp transition between the phases. 
Moreover, for $^{161}$Dy with $m_1=\sfrac{21}{2}$, $m_2=\sfrac{19}{2}$ we presented a nontrivial stability range due to the interplay between interactions. 
The generalization of our method may be adapted to other multicomponent phenomena in dipolar fermionic systems e.g the Einstein de-Haas effect or the spontaneous demagnetization.

\begin{acknowledgments}

We are grateful to A. Griesmaier and B.-G. Englert for helpful discussions.
P.B., K.P., and K.R. acknowledges support by Polish Government research grant N N202 174239 for the years 2010--2012. K.P., K.R., T.P. acknowledge financial support by contract research `Internationale Spitzenforschung II-2' of the Baden-W\"urttemberg Stiftung, ``Decoherence in long range interacting quantum systems and devices''. 
\end{acknowledgments}

\appendix
\section{Derivation of direct and exchange energy between two components}
The direct energy in \eqref{eqn:totalFun} between two components without correlations has the form:
\beal
E_{dir}^{12}=\frac{1}{2(2\pi)^6} \integral{^3\bx } \integral{^3\bk } \integral{^3\bx\rq{}} \integral{^3 \bk\rq{}} 
 \nn\\
\times f_{11}(\bx,\bk)f_{22}(\bx\rq{}) 
 V_{1122}(\bx-\bx\rq{}).\nn
\label{eqn:totalFun2ndDiag}
\end{align}
which after applying \eqref{eq:densReSpace} can be writen as:
\beq
E_{dir}^{12}=\frac{1}{2} \integral{^3\bx } \integral{^3\bx\rq{}} 
n_{11}(\bx,\bk) n_{12}(\bx\rq{}) 
V_{1122}(\bx-\bx\rq{}).\nn
\label{eqn:totalFun2ndDiag}
\eeq
Using Fourier transform of the  densities in position space: %(we replace double indices $ii$ by $i$):
\beq
\mathcal{F}\{n_{ii}\}(\bk)=N_{i}\exp\left[-\frac{k_x^2+k_y^2}{2 \beta_i \lambda_i \gamma^2} N_{i}^{1/3} -\frac{\beta_i^2k_z^2}{2 \lambda_{i} \gamma^2}N_i^{1/3}\right]\nn
\eeq
we get:
\beq
E_{dir}^{12}=\frac{1}{2(2\pi)^3} \integral{^3k}\mathcal{F}\{n_{11}\}(\bk)\mathcal{F}\{V_{1122}\}(\bk)\mathcal{F}\{n_{22}\}(\bk),\nn
\eeq
where the Fourier transform of the two body interaction is 
\beq
\mathcal{F}\{V_{1212}\}(\bk) = -g_{d} (1-3\cos^2\theta).\nn
\eeq
The $k$-integration is then performed in spherical coordinates.
After substitution of $\cos\theta$ by $u$ and performing integrals over $k$ and $\phi$ we get \eqref{eq:2CompEDir}.
The exchange energy between two components from \eqref{eqn:totalFun} has the form:
\beal
E_{ex}^{12}=- \frac{1}{2(2\pi)^6}  \integral{^3\bR } \integral{^3\bk_1 } \integral{^3 \bk_2} 
f_{11}(\bR,\bk_1) \nn\\
\times f_{22}(\bR,\bk_2) 
\mathcal{F}\{V_{1122}\}(\bk_1-\bk_2).\nn
\end{align}
After performming substitutions: $ \bk=\bk_1-\bk_2$ and $\bK=\frac{\bk_1+\bk_2}{2}$ we get:
\beal
E_{ex}^{12}=- \frac{1}{2(2\pi)^6}  \integral{^3\bR } \integral{^3\bk } \integral{^3 \bK} 
f_{11}(\bR,\bK+\frac\bk 2)\nn\\
\times f_{22}(\bR,\bK - \frac \bk 2) 
\mathcal{F}\{V_{1122}\}(\bk).\nn
\end{align}
We see from the form of the $f_{ii}$ Eq. \eqref{eqn:wignerGauss} that integral over $\bR$ can be performed independently as a Gaussian integral. While $\mathcal{F}\{V_{1212}\}(\bk)$ does not depend on $\bK$. Next we can easily calculate the Gaussian integral over $\bK$ in Cartesian coordinates. Finally, we transform the expression to spherical coordinates and after substituting $\cos \theta$ by $u$ we get Eq. \eqref{eq:2CompEEx}.
\section{Multicomponent noninteracting gas}
For multicomponent fermionic gas without dipolar and contact interaction in external $\bB$ field calculus of variation can be used to find out distribution of atoms between components and the shapes in momentum and position spaces.
The energy functional can be written using densities:
\beal 
\sum_{m}\bigg[\integral{^3x}\frac{3 \hbar^2}{10 M}(6 \pi)^{2/3}{ n_m(\bx)^{5/3}}
+\frac{M\omega^2 x^2}{2}n_m(\bx) \nn\\
-\Gamma n_m(\bx)  
-   {\mu_m} B n_m(\bx)\bigg],\nn
\end{align}
where $\Gamma$ is a Lagrange multiplier and we sum over components $m$.
\begin{figure}[h!]
\subfigure[ size in position space of each component]{\includegraphics[width=0.23\textwidth]{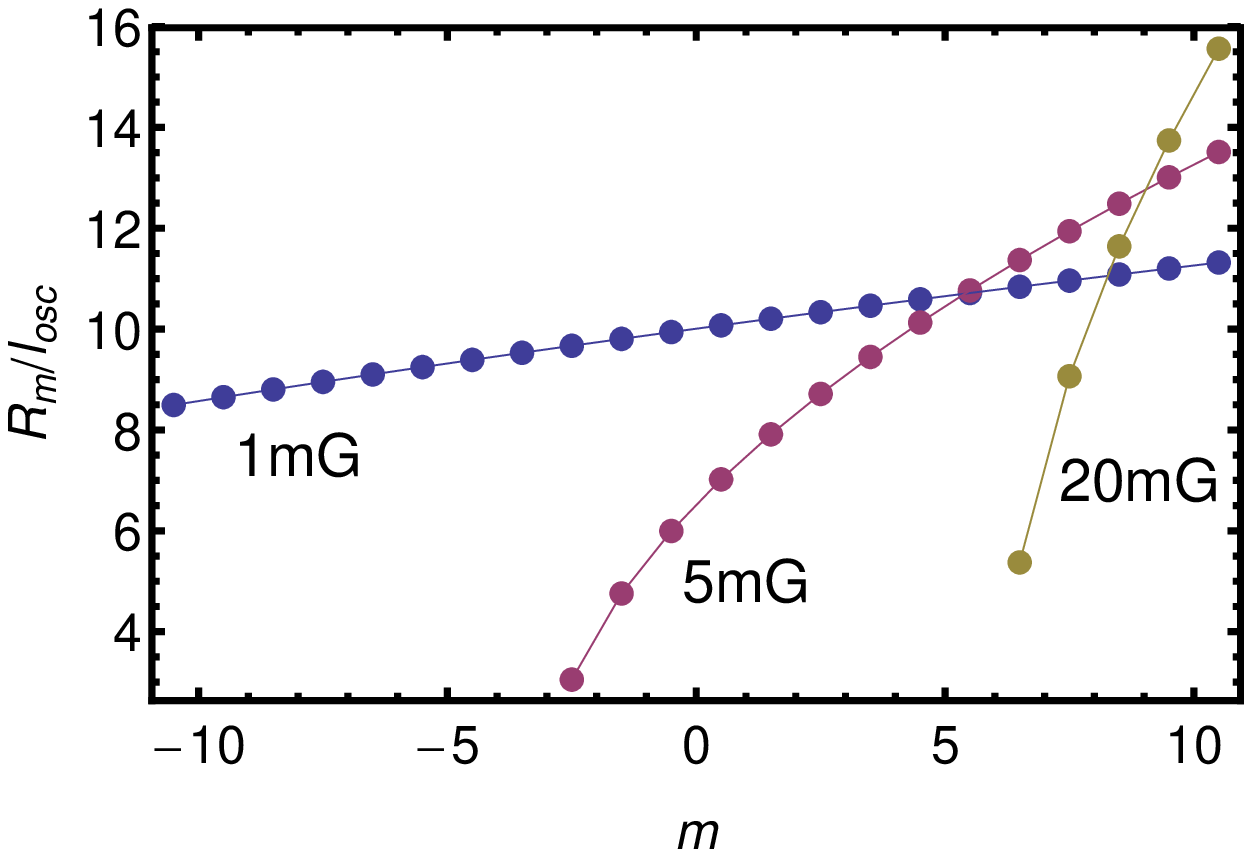}}
\subfigure[ number of atoms in each component]{\includegraphics[width=0.24\textwidth]{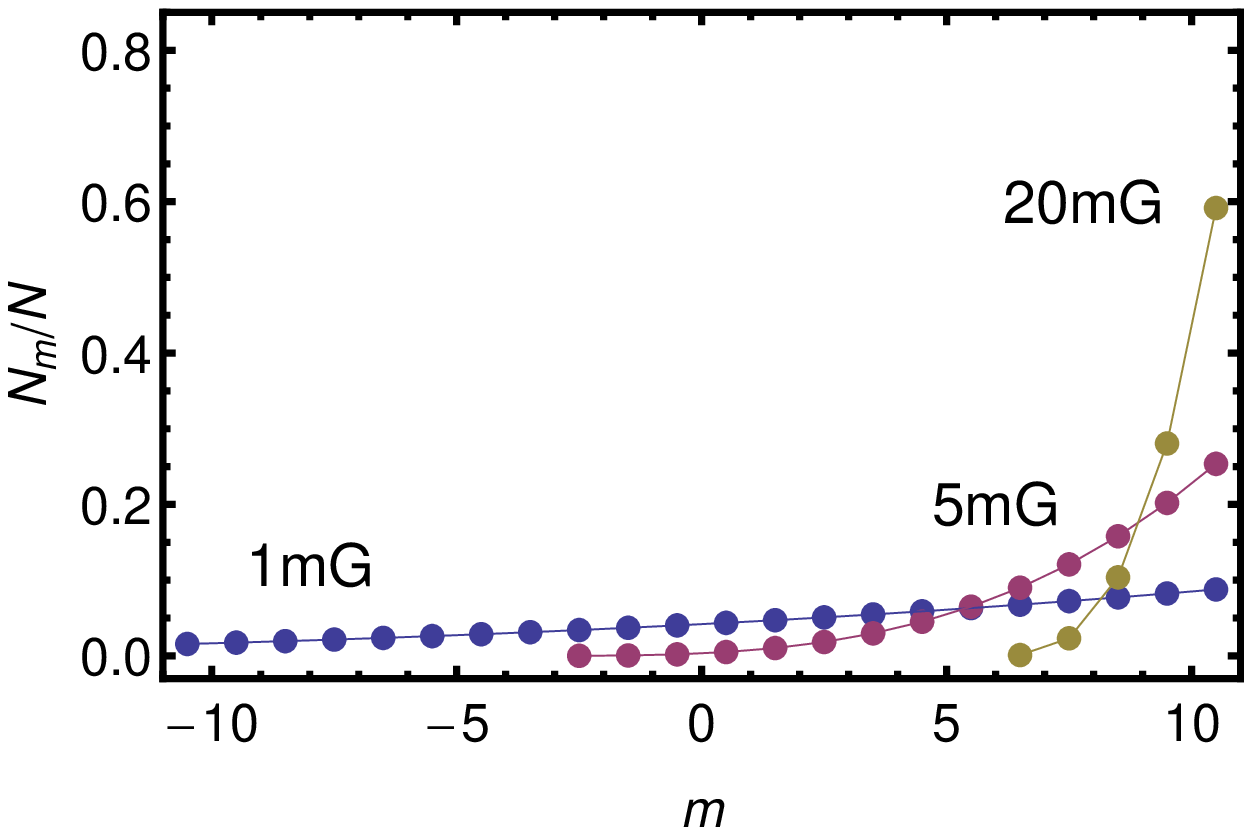}}
\caption{
 Size of the cloud and number of particles in every component for different magnetic fields. Results are for experimental parameters of $Dy$: $\mu=10 \mu_B$, $\omega=2\pi\times 100 Hz$ and $N=10^6$. 
}
\label{fig:sizeNum}
\end{figure}

Next, we vary the energy functional with respect to densities $n_m(\bx)$ getting for every $m$ the equation:
\beqs
\frac{5}{3}\frac{3 \hbar^2}{10 M}(6 \pi)^{2/3}{ n_m(\bx)^{2/3}}
+M\omega^2r^2/2
-\Gamma  
-   {\mu_m} B =0,
\eeqs
from which we can find the distribution in position space for every $m$ as a function of $\Gamma$:
\beq
n_m(\bx)=\frac{ (2 M)^{3/2}}{6 \pi^2 \hbar^3}[\Gamma 
+ \mu_m B 
- M \omega^2 x^2 /2]^{3/2}.\nn
\eeq
From the constraint on the total number of particles
\beqs 
N = \sum_{m}\integral{^3x}{n_{m}(\bx)} =
\sum_{m} \frac{M^{3/2} (\Gamma +\mu_m B)^3}{3 \left(M \omega ^2\right)^{3/2} \hbar ^3},\nn
\eeqs 
 we find $\Gamma$. Once we have $\Gamma$ we can calculate the number of atoms in every component and the size of the components in position space $R_m$ (radius for which $n(R_m)=0$).

Experimentally relevant $^{161}$Dy has magnetic dipole moment $\mu=10 \mu_B$  with 22 hyperfine levels with $m_F\in \{-21/2,\dots, 21/2\}$ and corresponding to it $\mu_m=\frac{20}{21} m_F \mu_B$.

Fig.~\ref{fig:sizeNum} presents the size of components of $^{161}$Dy as a function of $m_F$ for different values of the $\bB$ and $\omega=2\pi \times100$Hz and $N=10^6$.
The precision of $\bB$ field control (up to $10\mu G$ \cite{Pasquiou2011b}) is high enough to observe a true multicomponent ground state of the fermionic system with free magnetization in analogy to experiment with bosonic Cr \cite{Pasquiou2011b}.  
Such a thermalization to multicomponent state is not possible in the system with only contact interaction.

\bibliography{/Users/admin/Dropbox/notes/latex/library}
\end{document}